# First-Order Phase Transitions in a Quantum Hall Ferromagnet

Vincenzo Piazza[*], Vittorio Pellegrini[*], Fabio Beltram[*], Werner Wegscheider[†], Tomáš Jungwirth[‡,§], and Allan H. MacDonald[‡]

[*] *Scuola Normale Superiore and Instituto Nazionale per la Fisica della Materia, Pisa, Italy*

[†] *Walter Schottky Institute, Munich, Germany*

[‡] *Department of Physics, Indiana University, Bloomington, Indiana 47405*

[§] *Institute of Physics ASCR, Cukrovarnická 10, 162 00 Praha 6, Czech Republic*

**The single-particle energy spectrum of a two-dimensional electron gas in a perpendicular magnetic field consists of equally-spaced spin-split Landau levels, whose degeneracy is proportional to the magnetic field strength. At integer and particular fractional ratios between the number of electrons and the degeneracy of a Landau level (*filling factors* ν) quantum Hall effects[1] occur, characterised by a vanishingly small longitudinal resistance and quantised Hall voltage[2]. The quantum Hall regime offers unique possibilities for the study of cooperative phenomena in many-particle systems under well-controlled conditions. Among the fields that benefit from quantum-Hall studies is magnetism, which remains poorly understood in conventional material. Both isotropic and anisotropic ferromagnetic ground states have been predicted[3,4,5,6,7,8] and few of them have been experimentally studied[9,10,11,12,13] in quantum Hall samples with different geometries and filling factors. Here we present evidence of first-order phase transitions in $\nu = 2$ and 4 quantum Hall states confined to a wide gallium arsenide quantum well. The observed hysteretic behaviour and anomalous temperature dependence in the longitudinal resistivity indicate the occurrence of a transition between the two distinct ground states of an Ising quantum-Hall ferromagnet. Detailed many-body calculations allowed the identification of the microscopic origin of the anisotropy field.**

The study of quantum phase transitions[14,15], tuned by external fields rather than temperature, has enlarged the domain of phase-transition physics. In electronic systems, transitions to ordered ground states can occur when interactions between particles, rather than single-particle potentials, play a dominant role in selecting the ground state. This situation is frequently encountered in the quantum Hall regime owing to the quantisation of particle in-plane kinetic energy into macroscopically degenerate Landau levels. The ordered many-particle ground states discussed in this paper originates when two or, in general, $N$ Landau levels are brought close to alignment and the filling factor of these levels is an integer less than $N$[3,6,7,8]. The close analogy between these states and those of two-dimensional ferromagnets is often emphasised by using a pseudospin language[16] to describe the Landau level degree of freedom. For a two-level case, we refer to one of the single-particle Landau levels as the pseudospin-up state and to the other as the pseudospin-down state. A general quantum spinor with an arbitrary orientation is a linear combination of the up and down states.

Ferromagnetic states are characterised by pseudospin polarisation of the two-component many-particle system which remains finite even when the splitting between the two Landau levels is set to zero. In single-layer two-dimensional systems at $n = 1$ the pseudospin degree of freedom can be the real electron spin. In this case it is rigorously established[3] that the ground state is an isotropic strong ferromagnet. In general, the pseudospin degree of freedom can involve real-spin, orbit-radius quantum number, layer index in multiple quantum wells or subband index in wide quantum wells[5].

Electron-electron interactions are then expected to lead to tunable pseudospin anisotropy and to a complex range of ordered states which can be explored by adjusting system parameters, often in the same physical sample. In a double-quantum-well structure, for example, Landau level crossings can be accompanied by the introduction of broken symmetry phase with easy-plane pseudospin anisotropy whose soft collective excitations and related quantum phase transitions were recently observed.[11,12,13]

The possibility of realising quantum Hall ferromagnets with the *easy-axis* (Ising) anisotropy, common in familiar magnetic systems, has also been raised[5]. Experiments[17] on several fractional quantum Hall effect (QHE) transitions suggested interpretations within this framework; however, no rigorous physical picture has been established to describe quantum Hall ferromagnets at fractional filling factors. The experiments reported here reveal intriguing phenomena arising when Landau levels with opposite spins and different subband indices are brought close to degeneracy by applying an external electric field. The observed suppression of the $n = 2$ and 4 quantum Hall effects and hysteretic behaviour in the low-temperature longitudinal resistance indicate an undergoing first-order phase transition between oppositely polarised ground states of an Ising ferromagnet with domain structure. Detailed many-body calculations confirm that at external fields consistent with experiment the ground state of the system is a quantum Hall ferromagnet with easy-axis anisotropy.

The sample we measured consists of a 60 nm wide gallium arsenide (GaAs) quantum well positioned 75 nm below the surface and embedded between two hard-wall $Al_{0.25}Ga_{0.75}As$ barriers. An important feature of our system is a *soft* barrier originating from Coulomb interactions among electrons in the quantum well. This barrier separates the electronic system into two well defined, yet strongly coupled, layers (see the insets of Figs. 2a, 2c). We argue below that the soft barrier is the source of the anisotropy field leading to Ising ferromagnetism in our sample. The two lowest energy-subbands of the unbiased quantum well are occupied by electrons. When the perpendicular magnetic field is applied, these subbands form two distinct Landau level ladders, each consisting of alternating spin up ($\uparrow$) and spin down ($\downarrow$) Landau levels labelled by the orbit-radius quantum number *n*. A gate contact evaporated on the surface of the sample allows us to apply an electric field across the quantum well. The external electric field changes the two-dimensional electronic density in the well and, more importantly, varies the separation between the two lowest energy levels in the well. At bias potential $V_g$ = 0, the electron density is $5.5\times10^{11}$ cm$^{-2}$ and the mobility $2.6\times10^{6}\ \mathrm{cm}^2/\mathrm{Vs}$. In our experiments, we use the external bias potential to align different pairs of Landau levels.

Colour plots in Figs. 1a and 1b show the measured longitudinal resistivity $r_{xx}$ as a function of magnetic field *B* and applied bias, at temperature $T = 330\,\mathrm{mK}$; dark (bright) regions correspond to low (high) resistivity. At a given gate voltage, the minima of $r_{xx}$ as a function of *B* correspond to integer filling factors. The evolution of the minima can be traced in the figure by following dotted lines, labelled by the corresponding Landau level filling factor. For the gate bias values of Fig. 1 ($V_g > -0.26\,\mathrm{V}$), both Landau level ladders are populated and $r_{xx}$ exhibits a complicated pattern reflecting the crossing of Landau levels from different ladders.

In order to study phase transitions in our system we now focus on the even-integer filling factors $n = 2$ and $n = 4$. The evolution of the $n = 4$ state as the gate voltage is swept from $V_g = -0.06\,\mathrm{V}$ to $V_g = -0.19\,\mathrm{V}$ is shown in Fig. 2d. The quantum Hall minimum in $r_{xx}$ disappears near the predicted critical biases at which a filled Landau level crosses with an empty level (see caption of Fig. 2). Significantly, the suppression of the quantum Hall minimum in the longitudinal resistivity correlates with the emergence of hysteresis in $r_{xx}$ in up and down sweeps of the magnetic field. Representative curves are shown in Fig. 3a for $T = 330\,\mathrm{mK}$. The Figure indicates that the detailed shape and position of the observed hysteresis is sensitive to small changes in the electron density which follow each thermal cycling of the sample; differences in the observed critical bias potentials at which the QHE is suppressed and hysteresis emerges correspond to the density variation in different measurement sessions.

Figure 3b shows an enlarged view of the peak near to the $n = 4$ minimum of $r_{xx}$; here the hysteresis extends over a large region of magnetic fields between the $n = 4$ and $n = 3$ state and occurs at $V_g$ = -0.152 V. The temperature evolution of the hysteresis, presented in Fig. 3c, confirms that the characteristic energy scale involved in this phenomenon is that of the electron-electron interaction, since the hysteresis disappears at temperatures above the estimated[5] Ising critical temperature for a quantum Hall ferromagnet, $T_C \approx 1\,\mathrm{K}$. Thermal cycling of the sample was found to have a negligible effect on measured $T_c$.

A similar phenomenology applies to the $n = 2$ quantum Hall state evolution, shown in Fig. 2c. As in the $n = 4$ case, the onset of hysteresis (see the inset of Fig. 2c) and the suppression of the quantum Hall state are strictly correlated. The transport anomalies described above have not been observed at odd-integer filling factors. We attribute the appearance of hysteresis to the development of pseudospin magnetic order in the two-dimensional Ising universality class and explain below why, in our sample, easy-axis pseudospin anisotropy is expected for even-integer filling factors.

We recall that for a disorder free system at $T = 0$, easy-axis pseudospin anisotropy implies that external fields exceeding a coercivity field have to be applied in the direction opposite to the pseudospin orientation in order to produce the pseudospin reversal. This property leads to hysteretic behaviour of the pseudospin orientation and first order phase transitions at the extrema of the hysteresis loop. In real systems it is expected that disorder will produce a random effective external field, leading to the formation of domains with particular pseudospin orientations. Dissipation due to mobile charges created in domain walls, where the quasiparticle excitation gap is reduced, can lead to a breakdown of the quantum Hall effect as shown in Figs. 1, 2. In the disorder free limit, long-range order is destroyed at a critical temperature yielding a finite temperature continuous phase transition in the Ising universality class.

The hysteresis and anomalous temperature dependence of the $n = 2$ and $n = 4$ $r_{xx}$ traces presented above, fall into the class of effects expected for easy-axis ferromagnets with a domain structure. The knowledge of the physical meaning of the pseudospin degree of freedom allows to identify the possible microscopic origin of the easy-axis anisotropy in our sample. Assuming a many-body fermionic wavefunction constructed from one-particle states with common pseudospin orientation $q$ we calculated the expectation value of the many-body Hamiltonian ($q = 0$ corresponds to pseudospin up and $q = p$ to pseudospin down.) We focus on the simpler case of $n = 2$. We consider first a model, consisting of two infinitely narrow quantum wells separated by a rigid tunnelling barrier. Within this model the pseudospin dependent contribution to the ground state energy at small $V_g$ is given by[5,8] $E_q = (\Delta_Z - \Delta_T - V_g^2 / \Delta_T)\cos q$ and the critical gate potentials by $V^{c(1)} = -V^{c(2)} = -\sqrt{\Delta_Z \Delta_T - \Delta_T^2}$. In this expression $\Delta_T$ is the tunnelling gap, i.e., the energy splitting between the two lowest quantum levels in the unbiased double-well, and $\Delta_Z$ is the Zeeman splitting. At this level of approximation the pseudospin ferromagnet is isotropic since there is no preferred pseudospin orientation at $V_g = V^c$: $E_q(V_g = V^c) \equiv 0$. To give rise to easy-axis anisotropy, we must take into account the softness of the barrier separating the two layers. At, e.g., negative bias the $n = 0,\uparrow, L$ Landau level ($R$ and $L$ labels the ladder whose wavefunction has dominant weight near right(left)-hand side of the quantum well, respectively) is always occupied and thus do not contribute. The $n = 0,\downarrow, L$ and $n = 0,\uparrow, R$ are the two pseudospin levels. We find in our LDA calculations that at sufficiently high electron densities, consistent with the experimental density at which the $n = 2$ transition occurs, the tunnelling barrier is reduced when the $n = 0,\downarrow, L$ level is occupied compared to the case of occupied $n = 0,\uparrow, R$ Landau level. In the LDA calculations this occurs because the exchange energy which counters the electrostatic barrier between layers is larger in the latter, fully spin-polarised, case. From a many-body point of view, the reduction in the quasiparticle tunnelling gap reflects the correlated mixing of higher electronic subbands into many-body states. Smaller barrier leads to a larger tunnelling gap, i.e., to a larger separation between $L$ and $R$ Landau levels. Translated into the pseudospin language, the softness of the barrier leads to the increase of the gap between the occupied lower-energy pseudospin level and the empty higher-energy pseudospin level in either up or down pseudospin polarisations. We can account for this property by replacing $\Delta_T$, in the ground state energy expression given above, by $\Delta_T^0 + d\Delta_T \cos q$, where $d\Delta_T > 0$. The anisotropy energy term $-d\Delta_T \cos^2 q$ favours the pseudospin angle $q = 0$ or $p$, i.e., the system behaves like an Ising ferromagnet. Similar arguments explain easy-axis anisotropy at filling factor $n = 4$. For odd-integer filling factors the two pseudospin Landau levels have the same real-spin and the above scenario does not apply. The odd-integer quantum Hall effect states have easy-plane anisotropy consistent with our experimental observation.

These unusual magneto-transport observations are but one instance of a broad class of phenomena that can be associated with crossing Landau levels. Even for the relatively simple case of two crossing levels, broken-symmetry ground states analogous to those of isotropic, easy-plane or easy-axis two-dimensional ferromagnets can occur. The phenomenology of easy-axis systems, dominated by the hysteretic effects and domain-morphology issues familiar in common magnetic systems, is particularly rich. The observations presented here suggest that quantum Hall systems can be used as new ideal systems for the study of these intriguing phenomena.

Fig. 1 V. Piazza

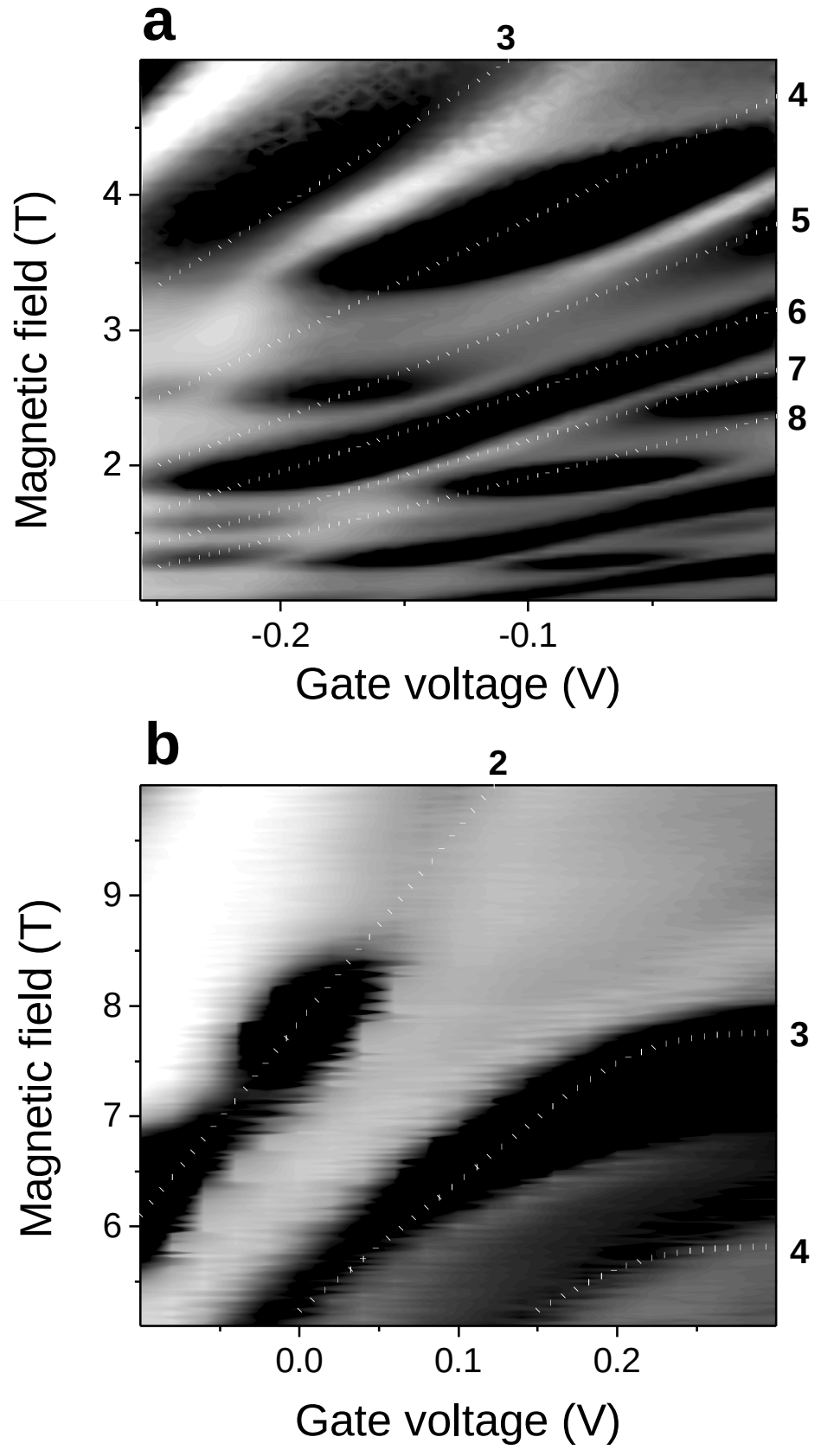

FIG. 1. Diagonal resistivity as a function of the magnetic field and gate voltage. Dark (bright) regions represent low (high) resistivity values. The dotted lines show the evolution of the quantum-Hall-effect minima and are labelled by the corresponding filling factors *n*. For gate voltages $V_g > -0.26\,\mathrm{V}$ two subbands are occupied: the complex resistivity pattern in this region originates from crossings between Landau levels of the two subbands. Similar evolution pattern of the resistivity is obtained in different measurement sessions after thermally cycling the sample. Changes of gate voltages of approximately 20 mV are observed in different measurement sessions owing to small differences in the total electron density. Measurements shown in all panels and subsequent figures were performed at 330 mK (unless stated) with lock-in techniques at a frequency of 12.5 Hz; the AC excitation current was set to 100 nA to avoid electron heating.

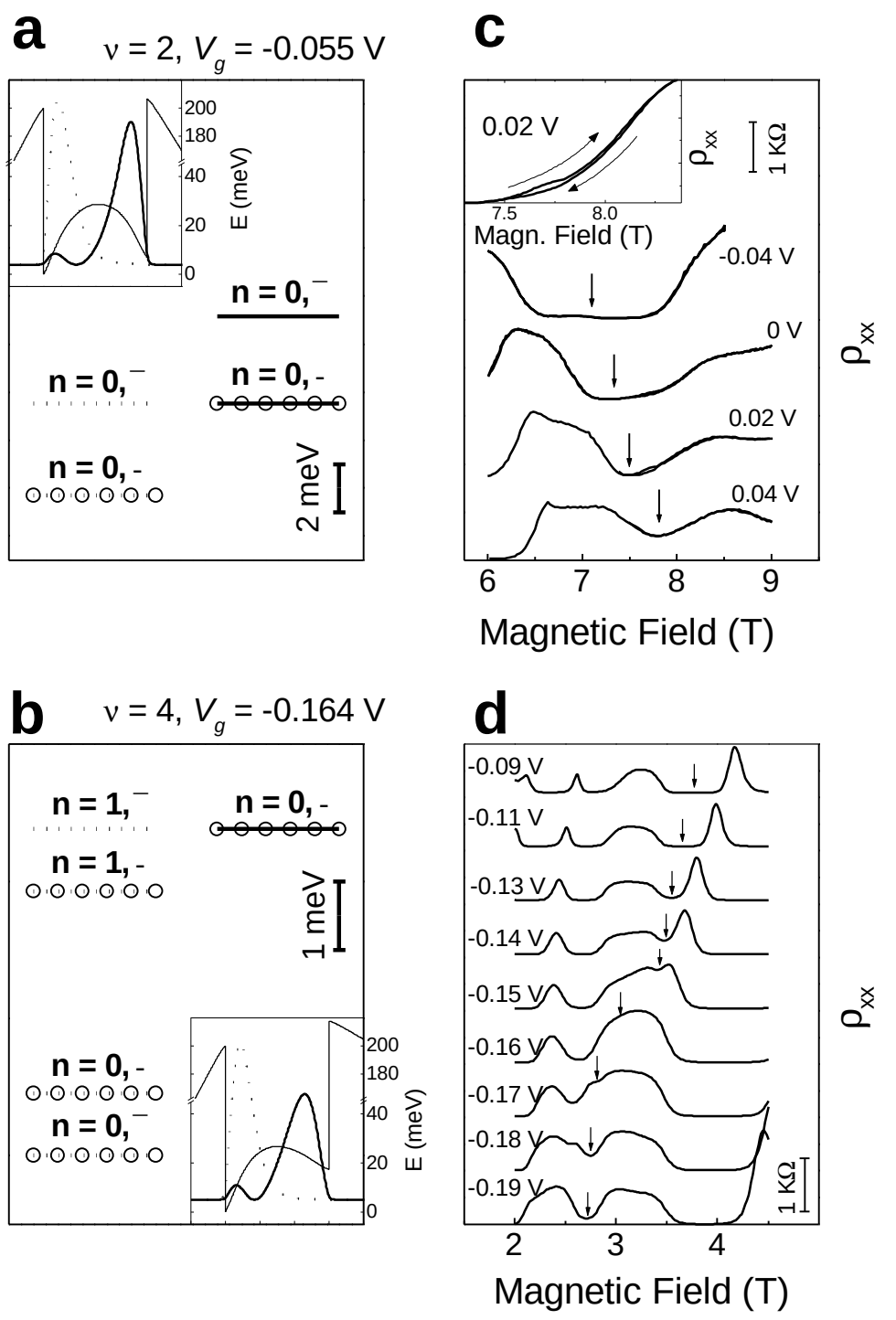

FIG. 2. (a, b) Self-consistent local-density-approximation (LDA) calculations of the energy spectrum for $n = 2$ (a) and $n = 4$ (b). Circles mark the occupied states. Insets show LDA quantum-well profiles and wavefunctions. Landau levels corresponding to states that occupy dominantly the right (left)-hand side of the well are labelled by index R (L). For $n = 2$, the filled $n = 0, \uparrow, R$ level ($n$ is the orbit-radius quantum number and $\uparrow$, $\downarrow$ label spin up, down) is brought to degeneracy with the empty $n = 0, \downarrow, L$ level at a critical potential $V_{n=2}^{c(1)} = -0.055\,\mathrm{V}$. Second critical potential $V_{n=2}^{c(2)} = -0.01\,\mathrm{V}$ corresponds to the case (not shown in figure) of aligned filled $n = 0, \uparrow, L$ and empty $n = 0, \downarrow, R$ Landau levels. At $n = 4$ the critical bias potentials are $V_{n=4}^{c(1)} = -0.164\,\mathrm{V}$ and $V_{n=4}^{c(2)} = -0.144\,\mathrm{V}$. We note that the LDA Landau level energies depend on which Landau levels are assumed to be occupied. For example at $n = 2$, the spin-polarised configuration with both R an L Landau ladders occupied and the spin-unpolarised case with electrons only in one of the ladders are both self-consistent LDA solutions in the interval between $V_{n=2}^{c(1)}$ and $V_{n=2}^{c(2)}$. This is consistent with the observed hysteretic behaviour. At $n = 4$, on the other hand, no self-consistent LDA ground state can be found when the bias is swept from $\approx -0.1\,\mathrm{V}$ to $V_{n=4}^{c(2)}$ or from $V_{n=4}^{c(1)}$ to $\approx -0.2\,\mathrm{V}$. Both bistability and failure to achieve self-consistency in the mean-field-like LDA calculations signal the importance of correlations in determining the many-particle ground state. (c, d) Set of resistivity-versus-magnetic-field curves recorded at different gate voltages showing the disappearance of the $n = 2$ (panel c) and $n = 4$ (panel d) quantum-Hall states (marked with arrows.) For the $n = 2$ case, the resistivity for up and down sweeps of the magnetic field is also shown in the inset for $V_g = 0.02\,\mathrm{V}$. Hysteretic behaviour appears when the gap responsible for the quantum-Hall state is strongly reduced. The curves are offset for clarity and each curve is labelled with the gate voltage.

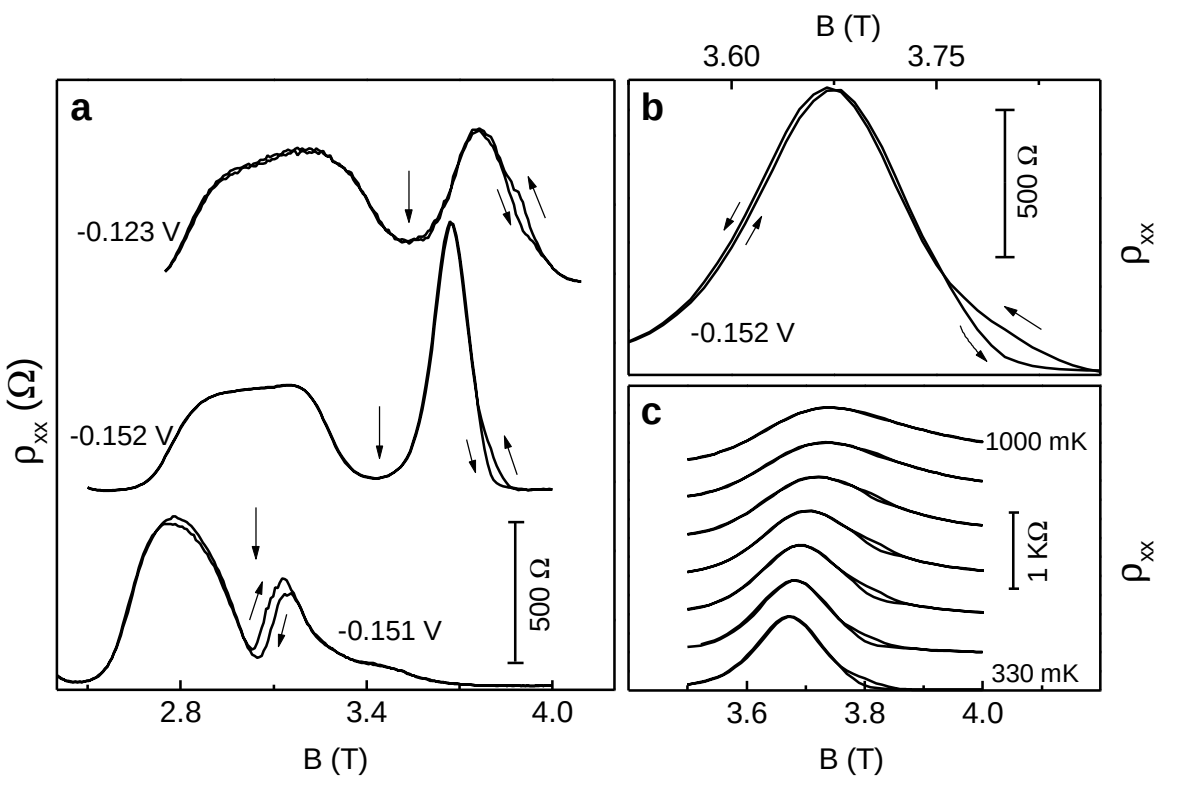

FIG. 3. (a) Diagonal-resistivity traces at 330 mK near filling factor $n = 4$ obtained after thermally cycling the sample for up and down sweeps of the magnetic field. The up and down sweeps are indicated by arrows. The presence of hysteresis is always correlated to the suppression of the $n = 4$ state (indicated by arrows). The shift in the position of the $n = 4$ state indicates small differences (less than 15%) in the total electron density obtained after thermally cycling the sample. (b) Enlarged view of the diagonal-resistivity trace shown in Panel (a) at $V_g$ = -0.152 V: the observed hysteresis extends over a large region of magnetic fields between $n = 4$ and $n = 3$. (c) Temperature evolution of the hysteresis at $V_g = -0.152$ V. The curves are offset for clarity.

[1] Prange, R. E. & Girvin, S. M., (eds), The Quantum Hall Effect. (Springer, New York, ed. 2, 1990).
[2] Das Sarma, S., & Pinczuk, A., (eds), Perspectives in Quantum Hall Effects. (Wiley, New York, 1996).
[3] Girvin, S. M. & MacDonald, A.H., in Ref. 2, pp. 161–224.
[4] Giuliani, G. F. & Quinn, J. J. Spin-polarization instability in a tilted magnetic field of a two-dimensional electron gas with filled Landau levels. *Phys. Rev. B* **31**, 6228–6232 (1985).
[5] Jungwirth, T., Shukla, S. P., Smrcka, L., Shayegan, M. & MacDonald, A. H. Magnetic anisotropy in quantum Hall ferromagnet. *Phys. Rev. Lett.* **81**, 2328–2331 (1998).
[6] Zheng, L., Radtke, R. J. & Das Sarma, S. Spin-excitation-instability-induced quantum phase transitions in double-layer quantum Hall systems. *Phys. Rev. Lett.* **78**, 2453–2456 (1997).
[7] Das Sarma, S., Sachdev, S. & Zheng, L. Double-layer quantum Hall antiferromagnetism at filling fraction $n=2/m$ where $m$ is an odd integer. *Phys. Rev. Lett.* **79**, 917–920 (1997).
[8] MacDonald, A. H., Rajaraman, R. & Jungwirth, T. Broken symmetry ground states in $n=2$ bilayer quantum Hall systems. *Phys. Rev. B* **60**, 8817–8826 (1999).
[9] Eisenstein, J. P., in Ref. 2, pp. 58–70.
[10] Daneshvar, A. J. *et al.* Magnetisation instability in a two-dimensional system. *Phys. Rev. Lett.* **79**, 4449–4452 (1997).
[11] Pellegrini, V. *et al.* Collapse of spin excitations in quantum Hall states of coupled electron double layers. *Phys. Rev. Lett.* **78**, 310–313 (1997).
[12] Pellegrini, V. *et al.* Evidence of soft-mode quantum phase transitions in electron double layers. *Science* **281**, 799–802 (1998).
[13] Sawada, A. *et al.* Phase transition in the $n=2$ bilayer quantum Hall state. *Phys. Rev. Lett.* **80**, 4534–4537 (1998).
[14] Sachdev, S. Quantum Phase Transitions. (Cambridge Press, Cambridge, 1999).
[15] Sondhi, S. L., Girvin, S. M., Carini, J. P. & Shahar, D. Continuous quantum phase transitions. *Rev. Mod. Phys.* **69**, 315–333 (1997).
[16] MacDonald, A. H., Platzman, P. M. & Boebinger, G. S. Collapse of integer Hall gaps in a double-quantum-well system. *Phys. Rev. Lett.* **65**, 775–778 (1990).
[17] Cho, H. *et al.* Hysteresis and spin transitions in the fractional quantum Hall effect. *Phys. Rev. Lett.* **81**, 2522–2525 (1998).

**Acknowledgements**. We thank M. Bichler for technical support during sample growth, and G.F. Giuliani for useful discussions. The work at Scuola Normale Superiore was funded in part by MURST. The work at Indiana University was supported by NSF grants, and at Intitute of physics ASCR by the Ministry of Education of the Czech Republic, and by the Grant Agency of the Czech Republic.